\documentclass[12pt,a4paper,twocolumn,fleqn]{narms}

\usepackage{graphicx,keyval,trig,url} 
\usepackage[T1]{fontenc}     
\usepackage[utf8]{inputenc}	
\usepackage{graphicx}

\usepackage{amsthm}
\usepackage{amssymb}
\usepackage{amsmath}
\usepackage{mathtools}
\usepackage[mathscr]{eucal}
\usepackage{lipsum}
\usepackage{cuted} 
\usepackage{blindtext}
\usepackage{color}
\usepackage{dsfont}

\usepackage{epstopdf}
\usepackage{bm}   

\usepackage{color}
\usepackage{xcolor}

\usepackage{multirow}
\usepackage{longtable}

\usepackage{mathtools}
\usepackage[colorinlistoftodos]{todonotes}

\usepackage{algorithm}
\usepackage{algpseudocode}

\usepackage{xcolor}
\usepackage{import}

\usepackage[normalem]{ulem} 

\usepackage{subfigure}
\usepackage{epsfig}
\usepackage{timesmt}
\usepackage{todonotes}
\usepackage{gensymb}
\usepackage{siunitx}

\usepackage{makecell}
\usepackage{tablefootnote}

\usepackage{chicaco}

\begin{document}
\title{A control-oriented wind turbine dynamic simulation framework which resolves local atmospheric conditions}
\author{{Z. Feng \& R. Ferrari \& J.W. van.Wingerden} \\
{\aff{Delft Center of System and Control}} \\
{\aff{Delft University of Technology, Delft, Netherlands}} \\
\\
{\authornext{Y. Liu}}\\
{\aff{Electric Power Research Institute (Europe), Dublin, Ireland}}} 

\date{}

\abstract{
Wind turbines may experience local weather perturbation, which is not taken into account by
the commonly-used wind turbine simulation packages.
Without this information, it is extremely challenging to evaluate the controller performance with regard to the effect of the variation of local atmospheric conditions. 
On the other side, it is too late and costly to wait until field test time.
To fill this gap, in this paper, we develop a control-oriented turbine dynamic simulation framework to evaluate the controller performance considering the 
perturbation of local atmospheric conditions. 
This goal is achieved by integrating an internal wind turbine (IWT) model in the Weather Research and Forecasting (WRF) simulation tool. 
The proposed framework is implemented on a 5MW reference wind turbine,
where the effects of the local atmospheric conditions are illustrated. The controller performance results are compared with those derived from the Fatigue, Aerodynamics, Structures,and Turbulence (FAST) simulator as a validation.
Simulation results show that the proposed WRF-IWT model is able to capture the turbine dynamics and controller performance regarding the perturbation of the complex local atmospheric conditions. The proposed framework can be leveraged to assist in designing controllers, which is more applicable to real-world wind conditions.
}

\maketitle

\title{A control-oriented wind turbine dynamic simulation framework which resolves local atmospheric conditions}
\author{{Z. Feng} \\
{\aff{Delft Center of System and Control}} \\
{\aff{Delft University of Technology}} \\
\\
{\authornext{Y. Liu}}\\
{\aff{Electric Power Research Institute (Europe)}} \\
{\aff{Dublin, Ireland}}\\
\\
{\authornext{R. Ferrari}}\\
{\aff{Delft Center of System and Control}} \\
{\aff{Delft University of Technology}}\\
\\
{\authornext{J.W. van Wingerden}}\\
{\aff{Delft Center of System and Control}} \\
{\aff{Delft University of Technology}}}

\section{INTRODUCTION}

In past decades, wind energy has received an increasing attention in the international energy market. This motivates more research on developing efficient wind turbine controller to optimize power generation and alleviate structural loads. However, it remains a challenge to simulate the generated power and the structural loads in real-world conditions, due to the complex time-varying meteorological phenomena that are involved. In addition, there is a lack of simulation tools oriented to wind turbine control that can account for the local atmospheric conditions during the controller design process. Finally, waiting for field tests to evaluate the influence of such local conditions on controller performance is neither timely, nor economically convenient. Therefore, there is an urgent need to develop a control-oriented numerical approach to simulate the interaction of a wind turbine with local atmospheric effects.

The generalized actuator disk (GAD) and the generalized actuator line (GAL) are two widely used turbine models to compute the aerodynamic loads under a given incoming flow field~\shortcite{BoersmaWFTutorial}. To capture the aerodynamic performance, they are integrated with the flow model in control-oriented numerical simulation tools, such as the Simulator for Wind Farm Applications (SOWFA)~\shortcite{MatthewSOWFA} and the Flow Redirection and Induction in Steady-state (FLORIS)~\shortcite{GebraadFLRIS} numerical model.
Although very effective, these models usually assume that the vertical shape of the wind flow can be ideally approximated by the uniform, power law, logarithmic wind profile models or a given boundary condition~\shortcite{sharmaEvolutionFlowCharacteristics2018}. 
However, such an assumption is not always verified when a wind farm experiences short-term local weather perturbation~\shortcite{YLiu2017}.
This, due to the lack of the atmospheric information, will lead to the sub-optimal design of the control system of the wind turbine. 
Thus, the designed controller based on these simulation tools might be less applicable for the wind turbine operating in the real world~\shortcite{rai2017}.

To consider a condition which is more representative of actual ones, the wind field can be derived by a physics-based model. WRF is an advanced atmospheric prediction system that integrates fully compressible, non-hydrostatic Euler equations~\shortcite{skamarockDescriptionARW}. The solver of WRF has the ability to resolve the flow around the wind turbine in a realistic, time-varying atmospheric environment. For this reason the wind field solution produced by WRF is widely used in SOWFA as an initial condition~\shortcite{Churchfield2013,Mache2014}. A WRF based large eddy simulation (LES) model with control strategies was developed to consider the local atmospheric condition above complex terrains in ~\shortcite{wangMultiscaleNumericalFramework2020}. A control-oriented LES model was designed to simulate the wind turbine wake, considering the Coriolis force, in~\shortcite{qianControlorientedLargeEddy2022a}. Nevertheless, these numerical simulation methods use a one-directional coupling from the flow to the turbine, where the effects that turbines have on the surrounding wind field are not be considered. Thus, the finer details of the actual atmospheric conditions around turbines cannot be resolved.

To derive a more accurate atmospheric conditions, in previous literature a turbine simulator is developed by two-way coupling between a turbine model and WRF-GAD model is integrated in the WRF code~\shortcite{MirochaWRFGAD}. The proposed model is able to simulate the turbine aerodynamic performance in realistic atmospheric condition. Nevertheless, it simplifies the model of rotation speed of the turbine by a piece-wise linear function versus wind speed, where the generator dynamics and control law are not captured. Thus, it leads to a limited capability of serving for control-oriented simulation in realistic atmospheric conditions.
In addition, both the WRF-GAD model and the WRF-LES models mentioned above are not able to consider the turbine structural dynamics model. Thus, they are not able to simulate the turbine structural vibration and capture its effects, such as the influence of the relative wind speed at the rotor due to the tower movement.  

To fill this gap, in this paper we develop a control-oriented wind turbine dynamic simulation framework which can be used to validate a given control design under local atmospheric conditions. This goal is achieved by integrating an internal wind turbine (IWT) model in the Weather Research and Forecasting (WRF) simulation tool. 
WRF is a physics-based meteorological simulation tool, which is capable of generating a variety of local atmospheric characteristics by a series of parameterized physical schemes. Thus, it is able to simulate the perturbation of local atmospheric conditions. In addition, the IWT model includes a generalized actuator disk (GAD) model, drive-train dynamics and a tower top vibration model. It is then integrated in WRF such that the turbine and control dynamics under the time-varying local atmospheric conditions can be taken into account. 
We employ the proposed model to numerically simulate a 5MW wind turbine developed  by  the US  National  Renewable  Energy Laboratory (NREL)~\shortcite{jonkman2009}. The turbine is situated near the Oceanic Platform of the Canary Islands (PLOCAN) site. The effects of the local atmospheric conditions are illustrated on the 5MW turbine with a BTC. Simulation results show that the proposed WRF-IWT model is able to capture the turbine dynamics and controller performance regarding the perturbation of the complex local atmospheric conditions. The results of controller performance are compared those derived from the Fatigue, Aerodynamics, Structures, and Turbulence (FAST) simulator. The comparison validates the drive-drain dynamics and controller works properly in the proposed WRF-IWT model. The results show a significant potential for using the proposed framework to develop and validate wind turbine control law under time-varying realistic atmospheric conditions. 

The remainder of this paper is organized as follows: Section 2 outlines the theoretical framework of WRF and GAD models. Section 3 introduces the tower top vibration model, where its effects on the relative wind speed is elaborated. In Section 4, a case study is carried out to exhibit its potential for controller design and to verify the results. Finally, conclusions are drawn in Section 5.




\section{Numerical model}

This section introduces the theoretical framework of the GAD model developed in WRF, the drive-train dynamics and the BTC algorithms. 

\subsection{WRF-GAD Model}
\begin{figure}[htpb]
      \centering
      \includegraphics[width=\columnwidth]{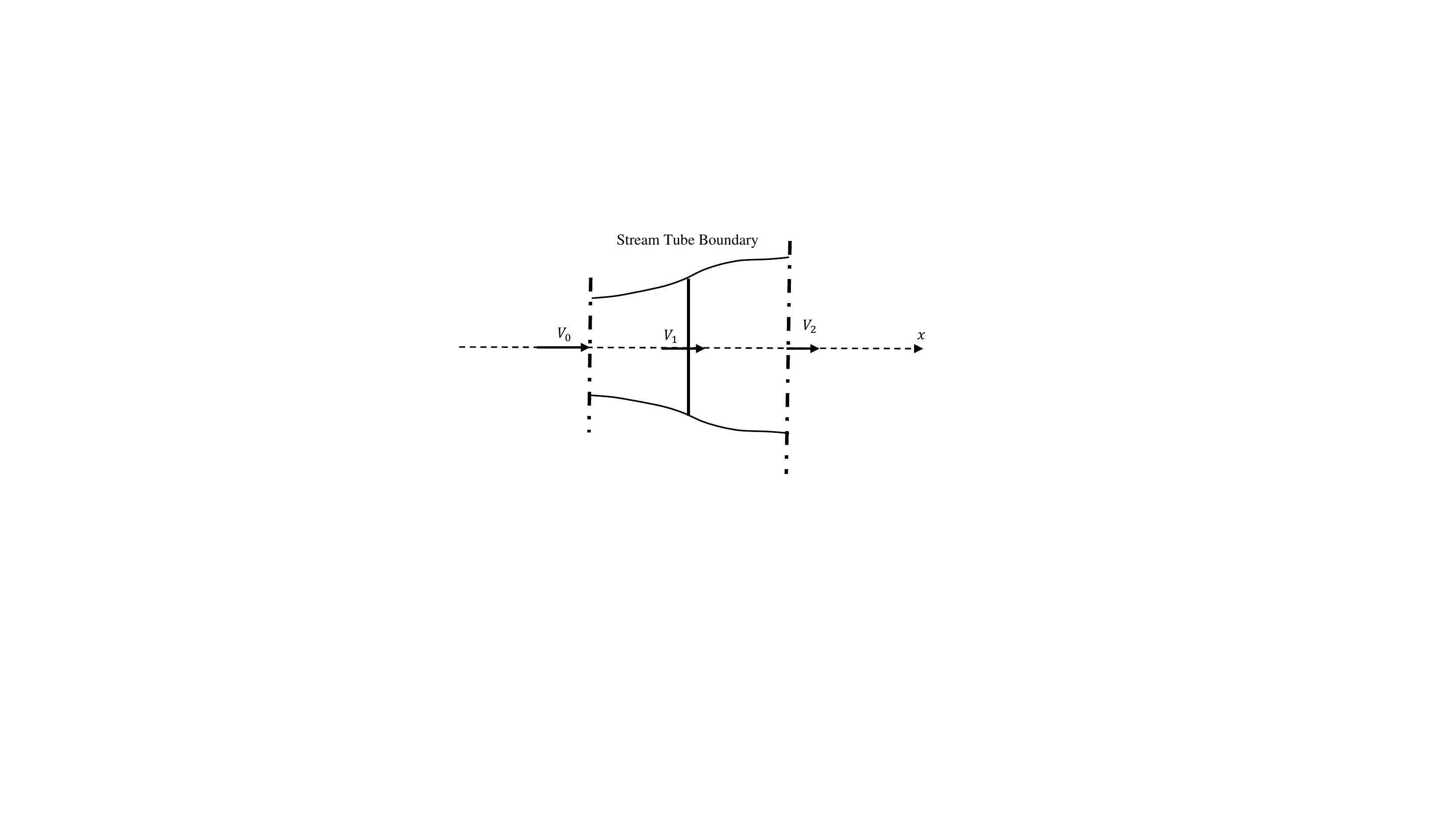}
      \caption{GAD model schematic.}
      \label{fig:GAD}
\end{figure}

The schematic of GAD model in presented in Fig.\ref{fig:GAD}, where $x$ is the streamwise direction aligned with the incoming wind speed $V_{\mathrm{0}}$. $V_{\mathrm{1}}$ is the velocity normal to the disk, which is reduced from its upstream value as a result of the widening of the stream tube. The induction factor $a_{\mathrm{n}}$ is introduced to account for this effect, then we have:
\begin{equation*}
    V_{\mathrm{1}} = V_{\mathrm{0}} (1-a_{\mathrm{n}})
    \, .
\end{equation*}
A similar expression can be derived for the downstream velocity $V_{\mathrm{2}}$ based on Bernoulli’s equation~\shortcite{burtonWindEnergyHandbook2011} as:
\begin{equation*}
    V_{\mathrm{2}} = V_{\mathrm{0}} (1-2a_{\mathrm{n}})
    \, .
\end{equation*}

\begin{figure}[thpb]
      \centering
    \includegraphics[width=0.9\columnwidth]{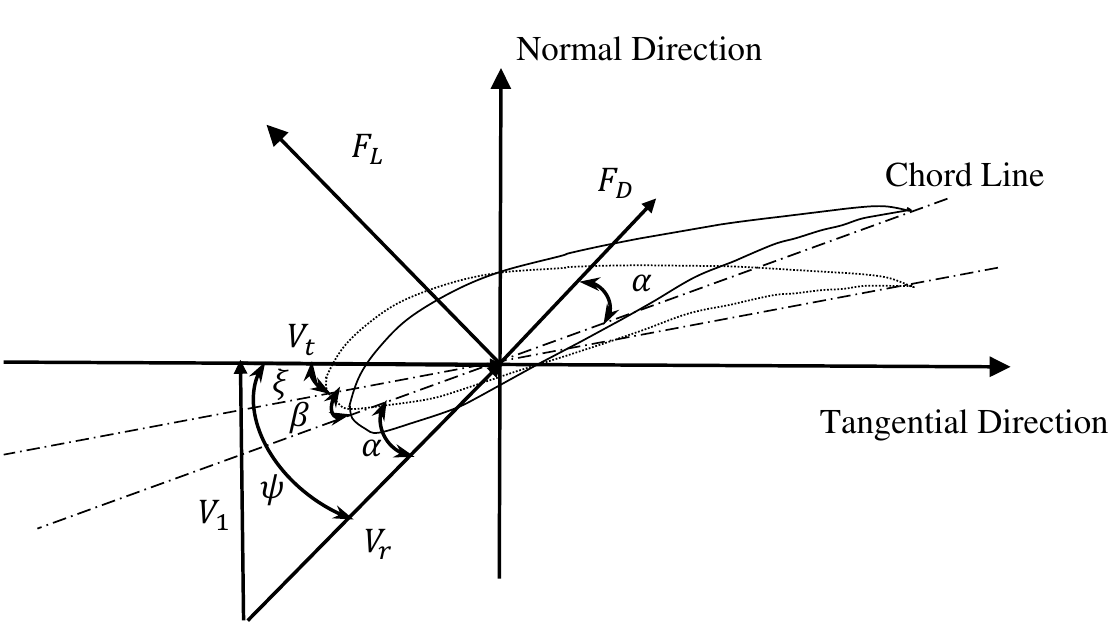}
      \caption{Forces and angles on an airfoil section, where $\alpha$ is the advance angle, $V_{\mathrm{r}}$ is the relative wind speed.}
      \label{fig:Forces and Angles}
\end{figure}

In GAD model, the turbine blades are represented by a rotating disk model, where the lift and drag forces are calculated based on the blade element theory~\shortcite{burtonWindEnergyHandbook2011}. The forces and angles on an airfoil section are presented in Fig.\ref{fig:Forces and Angles}. The relation of the angles then is derived by:
\begin{equation*}
    \beta = \psi - \alpha -\xi
    \, ,
\end{equation*}
where $\beta$ is the pitch angle, $\alpha$ is the angle of attack, $\psi$ is the advance angle, $\xi$ is the twist angle. The lift $F_{L}$ and drag $F_{D}$ produced by the wind can be derived \shortcite{manwellWindEnergyExplained2002} as:
\begin{equation}
\begin{aligned}
&F_{L}=\frac{1}{2} \rho V_{r}^{2} c C_{l} \\
&F_{D}=\frac{1}{2} \rho V_{r}^{2} c C_{d} 
\end{aligned}
\, ,
\label{eq:Lift and Drag}
\end{equation}
where $\rho$ is the air density, $V_{r}$ is the relative wind speed over the airfoil, $c$ is the the chord length of the blade, and $C_{l}$ and $C_{d}$ are the coefficients of lift and drag, respectively.

Then the axial force $ F_{\mathrm{n}}$ and the tangential force $ F_{\mathrm{t}}$ of each unit length can be derived by:
\begin{equation}
\begin{aligned}
& F_{\mathrm{n}}= F_{\mathrm{L}} \cos \psi+ F_{\mathrm{D}} \sin \psi \\
& F_{\mathrm{t}}= F_{\mathrm{L}} \sin \psi- F_{\mathrm{D}} \cos \psi
\end{aligned}
\, ,
\label{eq:FN FT}
\end{equation}

\begin{figure}
    \centering
    \includegraphics[width=\columnwidth]{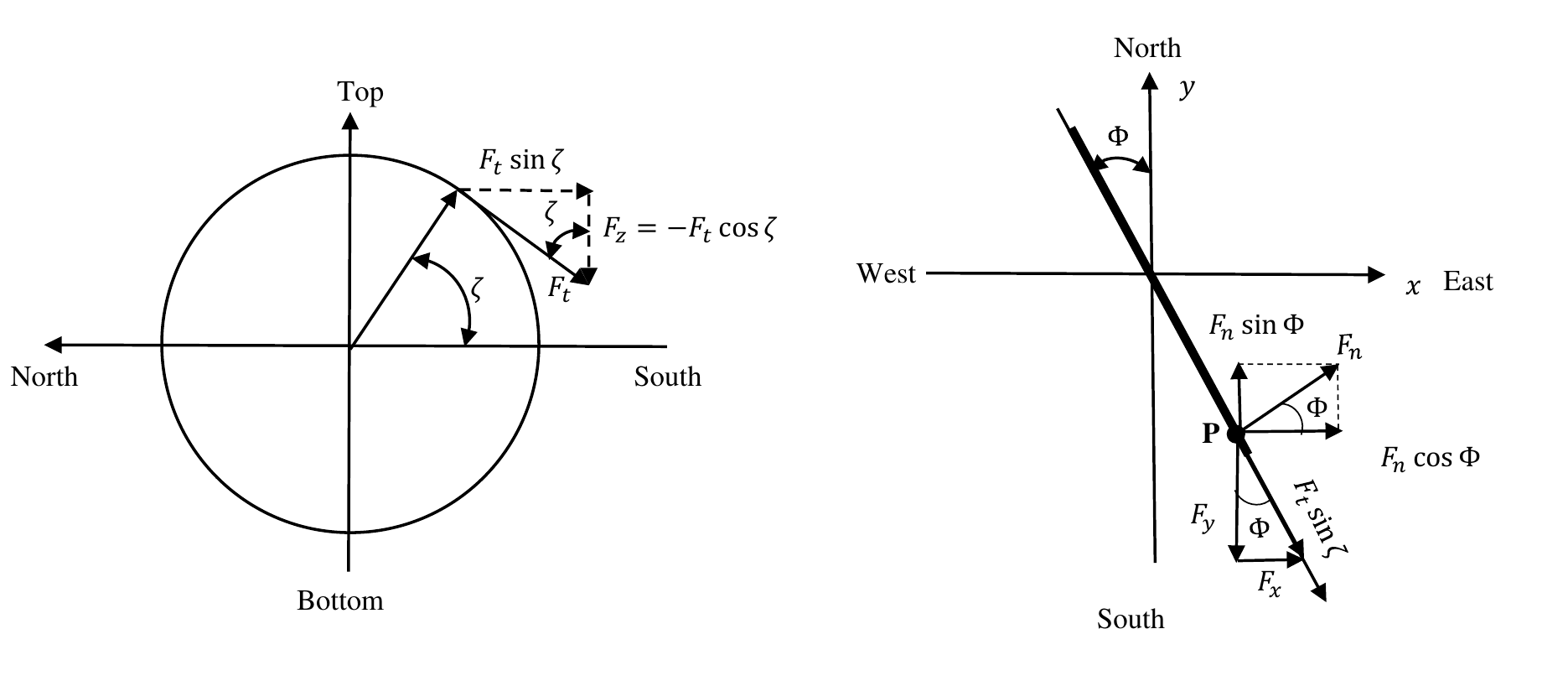}
    \caption{The projection of $\mathrm{d} F_{n}$ and $\mathrm{d} F_{t}$ onto the $\left[x,y,z \right]$ coordinate system in WRF}
    \label{fig:projection}
\end{figure}

The projection of $\mathrm{d} F_{n}$ and $\mathrm{d} F_{t}$ onto the $\left[x,y,z \right]$ coordinate system in presented in Fig.\ref{fig:projection}. The relations can be derived~\shortcite{MirochaWRFGAD} as:
\begin{equation}
\begin{aligned}
&F_{x} = F_{n} \cos \Phi+F_{t} \sin \zeta \sin \Phi \\
&F_{y}= F_{n} \sin \Phi-F_{t} \sin \zeta \cos \Phi \\
&F_{z}=-F_{t} \cos \zeta
\end{aligned}
\, ,
\end{equation}
where $\Phi$ and $\zeta$ describe the orientation of P on the actuator disk with respect to $\left[x,y,z \right]$ coordinate system in WRF. Then $\mathrm{F}_{x}$, $\mathrm{F}_{y}$ and $\mathrm{F}_{z}$ are added to the momentum equation in WRF~\shortcite{MirochaWRFGAD} as:
\begin{equation}
\begin{aligned}
&\frac{\partial u}{\partial t}=-\frac{1}{2 \pi r \rho} G\left(d_{n}\right) F_{x} \\
&\frac{\partial v}{\partial t}=-\frac{1}{2 \pi r \rho} G\left(d_{n}\right) F_{y} \\
&\frac{\partial w}{\partial t}=-\frac{1}{2 \pi r \rho} G\left(d_{n}\right) F_{z}
\, ,
\end{aligned}
\end{equation}
where $u$, $v$, and $w$ are the the three components of velocities. And $G\left(d_{n}\right)$ is a normal distribution, which is used to avoid numerical instabilities due to the strong forces applied at isolated grid points. $G\left(d_{n}\right)$ is with 0 mean and a standard standard deviation that is set to be the grid size.

Based on the above momentum analysis, $F_{\mathrm{t}}$ is derived by Eq.(\ref{eq:FN FT}). Then the aerodynamic power $P_{\mathrm{r}}$ is defined as:
\begin{equation}
    P_{\mathrm{r}} = \sum F_{\mathrm{t}}rdr
    \, .
    \label{eq:Pr}
\end{equation}
Then we can derived $T_{\mathrm{r}}$ by:
\begin{equation*}
    T_{\mathrm{r}} = P_{\mathrm{r}} / \omega_{\mathrm{r}}
    \, .
\end{equation*}

\subsection{Drive Train Dynamics}

For the WT dynamics, we model the turbine, drive-train and generator as a single rotational system, with generator speed ${\omega}_g$ in $rad/s$, and rotor speed ${\omega}_r(t)={\omega}_g/N$ in $rad/s$, where $N$ is the gear ratio of the drivetrain. We let $J_g$ and $J_r$ denote the inertia of the generator and rotor, respectively, and we let $J = J_g + J_r/N^{2}$ denote the equivalent inertia at the generator shaft. Neglecting losses, the dynamics is given by \shortcite{Hovgaard2015} as:
\begin{equation*}
J \dot{\omega}_{\mathrm{g}}(t)=T_{\mathrm{r}}(t) / N-T_{\mathrm{g}}(t)
\, ,
\label{DriveTrain}
\end{equation*}
where $T_{\mathrm{r}}(t)$ and $T_{\mathrm{g}}(t)$ are the rotor and generator torque, respectively.

\subsection{Baseline Torque Control Algorithms}
Then we develop the BTC based on the $K\omega^2$ torque control law~\shortcite{jonkman2009} as:
\begin{equation}
\tau_{\mathrm{g}}=\mathrm{K} \omega_{\mathrm{r}}^{2} / N
\, ,
\label{BTC}
\end{equation}
where $\tau_{\mathrm{g}}$ is the generator torque. $\mathrm{K} \in \mathbb{R}^{+}$ is the optimal mode gain derived by. 
\begin{equation*}
    \mathrm{K}=\frac{\pi \rho R^{5} \mathrm{C}_{\mathrm{p}}(\lambda, \beta)}{2 \lambda^{3}}
    \, ,
\end{equation*}
where $R$ is the rotor radius, $\beta$ is the blade pitch angle, $v_w$ is the wind speed, $\lambda$ is the tip-speed ratio. $C_p$ is the power coefficient, which can be derived from a lookup table.

\section{WRF-IWT MODEL}

In this section, the tower vibration model in WRF-IWT framework is introduced, as well as its influence on the relative wind speed between the rotor and the inflow wind.

\subsection{ Tower Top Vibration Model in WRF}
The thrust on an annular cross-section $\mathrm{d}T$ can be determined by the following expression that uses $a_{\mathrm{n}}$~\shortcite{MirochaWRFGAD,manwellWindEnergyExplained2002} as:
\begin{equation}
\mathrm{d}T= BF_{\mathrm{n}}\mathrm{d}r 
\, ,
\label{eq:Thrust}
\end{equation}
where $F_{\mathrm{n}}$ is derived by Eq. (\ref{eq:FN FT}), $B$ is the number of blades, which equals to 3. Then the total thrust force $T$ is calculated as:
\begin{equation}
   T = \sum \mathrm{d}T
    \, .
\end{equation}

The dynamics of the fore-aft bending mode of the tower is modeled as a second-order system as Eq.(\ref{eq:Tower}) \shortcite{Shaltout2017}:
\begin{equation}
M_{\mathrm{T}} \ddot{x}_{\mathrm{T}}+B_{\mathrm{T}} \dot{x}_{\mathrm{T}}+K_{\mathrm{T}} x_{\mathrm{T}}=T
\, ,
\label{eq:Tower}
\end{equation}
where $x_{\mathrm{T}}$ is the fore-aft displacement of the tower top, and $M_{\mathrm{T}}$, $B_{\mathrm{T}}$ and $K_{\mathrm{T}}$ are the tower equivalent mass, structural damping, and bending stiffness, respectively.

\subsection{ Interaction between Tower Vibration and Inflow Wind}
\begin{figure}
    \centering
    \includegraphics[width=0.7\columnwidth]{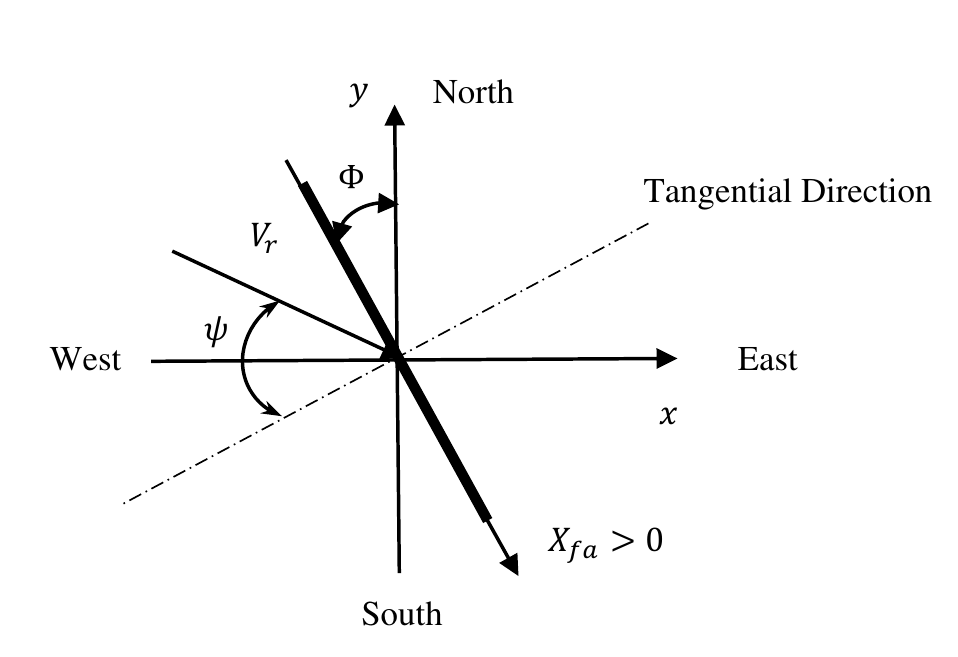}
    \caption{The relation between WRF and tower vibration coordination}
    \label{fig:ActualVr}
\end{figure}

To construct an accurate turbine simulation approach, it is important to consider the effects on relative wind speed due to the turbine vibration. As shown in Fig. \ref{fig:ActualVr}, the tower fore-after vibration $\vec{V}_{\mathrm{fa}}$ is along the axial of the turbine, the same direction as $\vec{V}_{\mathrm{r}}$. Then we can derive the actual relative wind speed $\vec{V}_{\mathrm{ac}}$ as:
 \begin{equation}
\vec{V}_{\mathrm{ac}}=\vec{V}_{\mathrm{r}} + \vec{V}_{\mathrm{fa}} 
\, .
\end{equation}
Then the $\vec{V}_{\mathrm{ac}}$ is used in Eq.(\ref{eq:FN FT}) to derive more accurate nominal and tangential forces. Then based on Eq.(\ref{eq:Pr}) and Eq.(\ref{eq:Thrust}), power and thrust force are derived.

Based on this WRF-IWT model, the proposed framework is able to take into consideration of the effect of the tower vibration on the aerodynamics loads and power production.

\section{SIMULATION}
In this section, the performance of the proposed numerical simulation framework is exhibited on the NREL 5MW reference wind turbine with BTC. In section A and B, the turbine parameters and WRF configuration are introduced, respectively. In section C, the results derived by the WRF-IWT framework are presented, which are used to verify the BTC performance under local atmospheric conditions near PLOCAN site. The results are also analyzed by being compared with those derived from FAST.

\subsection{Wind Turbine Configuration}
The wind turbine is situated in the PLOCAN site. The parameters of NREL-5MW wind turbine model are listed in Table \ref{ModelParameter}. The tower top vibration model is constructed based on the natural frequency $f_1$ and the damping ratio $d_1$ of the first fore-aft vibration mode, based on Eq.(\ref{eq:Tower}).

\begin{table*}[htpb]
\caption{ NREL-5 MW wind turbine model parameters }
\label{ModelParameter}
\begin{center}
\begin{tabular*}{\textwidth}{@{\extracolsep\fill}ll@{}}

\hline
Parameter & Magnitude \\
\hline
Rated power, $P_{\mathrm{g,rated}}$ & 5 MW\\
Rotor diameter, $D_\mathrm{r}$ & 126 m\\
Tower height, $H_\mathrm{t}$ & 87.5 m \\
Cut-in wind speed & 3 m/s \\
Cut-out wind speed & 25 m/s\\
Rated wind speed & 11.4 m/s\\
Rated generator speed, $\omega_{\mathrm{g,rated}}$ & 122.9096 rad/s\\
Minimum generator speed, $\omega_{\mathrm{g,}\min}$ & 41.6470 rad/s\\
Maximum generator speed, $\omega_{\mathrm{g,}\max}$ & 159.7824 rad/s \\
Maximum generator torque, $T_{\mathrm{g}}$ & $4.3094 \times 10^4$ N$\cdot$m \\
Generator  efficiency, $\eta$ & 0.9440 \\
Gear ratio, G & 97\\
Rotor Inertia, $J_\mathrm{r}$ & 35,444,067 kg/m$^2$ \\
Generator inertia, $J_\mathrm{g}$ &  534.116 kg/m$^2$ \\
Optimal tip-speed ratio, $\lambda^0$ & 7.6 \\
Optimal blade pitch angle, $\beta^0$ & 0 deg \\
Maximum power coefficient, $C_{\mathrm{p,}\max}$ & 0.4868 \\
Mass nacelle, $M_\mathrm{N}$ & 240000 kg \\
Mass hub, $M_\mathrm{H}$ & 56780 kg  \\
Mass blade, $M_\mathrm{B}$ & 17848.770 kg \\ 
1st tower fore-Aft natural frequency, $f_{1}$ & 0.3240 Hz\\
1st mode damping ratio, $d_1$ & 1\% \\
\hline
\end{tabular*}
\end{center}
\end{table*}

\subsection{WRF Configuration}
The wind turbine is situated in the PLOCAN site. The reanalysis data from the Global Forecast System (GFS)~\cite{cisl_rda_ds083.3} are used as initial/boundary conditions. They are derived from high-resolution forecasts in both spatial and temporal spaces. 

The resolution of the GFS dataset is on a $0.25 $ by $0.25 $ degree global latitude longitude grid, which is around $30 \times 30$ km. With the WRF Preprocessing System (WPS), the GFS input data can be interpolated to the horizontal grid spacing of $2250 \times 2250$ km. Then it is designed to be the first domain. A 5-domain nesting method is implemented to provide the grid scale fit for the individual turbine simulation. The details of the nesting method are presented in Table \ref{Table:GridDesign}. This nesting method is designed according to the nesting strategy and analysis in~\shortciteN{Charles2012}. The The wind turbine locates in the 5$^{\mathrm{th}}$ domain,where the horizontal grid size is set to be 10 meters. The first two domains are used for mesoscale simulations, while the finest three are used for the large-eddy simulation (LES) technique to simulate the turbine aerodynamics performance. Therefore, the Mellor-Yamada-Janjic (MYJ) scheme is selected as planetary Boundary layer (PBL) of the first 2 domains, while the LES is selected for the finer 3 domains. Here, MYJ scheme is an Eta operational scheme, which is a one-dimensional prognostic turbulent kinetic energy scheme with local vertical mixing~\shortcite{elbertWRFChemVersionUser2017}. The Grell scheme is only used in the coarsest domain because in a clear day the scheme is not relevant~\shortcite{Charles2012}. The Smagorinsky closure is chosen as it is recommended for real-data case~\shortcite{elbertWRFChemVersionUser2017}. More details of physics options used for the five domains are presented in Table \ref{Table:WRFPhy}. 

\begin{table*}[tbhp]
\caption{WRF Grids Design}
 \centering
\begin{tabular*}{\textwidth}{@{\extracolsep\fill}lllllll@{}}
\hline
 Domain & \makecell{PBL \\ scheme}  & \makecell{Land surface \\ model}  & Closure & \makecell{Cumulus \\Parameterization}  &    & Radiation \\
  \hline
1 & MYJ & NJSM & Smagorinsky & Grell scheme &  & RRTM \\
2 & MYJ  & NJSM & Smagorinsky & - &   & RRTM  \\
3 & LES & NJSM & Smagorinsky & - &  & RRTM  \\
4 & LES & NJSM & Smagorinsky & - &  & RRTM   \\
5 & LES & NJSM & Smagorinsky   & - &  & RRTM   \\
\hline
\multicolumn{7}{l}{\small *NJSM: Noah Land Surface Model; RRTM: Rapid Radiative Transfer Model.} \\
\multicolumn{7}{l}{\small *The Radiation includes both longwave and shortwave radiation.}
\end{tabular*}
\label{Table:WRFPhy}
\end{table*}

The simulations last for 180 seconds. The first 60s are truncated to avoid the instability at the beginning periods. The output data from WRF is derived every second.

\begin{table*}[tbhp]
\caption{WRF Grids Design}
 \centering
\begin{tabular*}{\textwidth}{@{\extracolsep\fill}llllllllll@{}}
\hline
  & nx  & ny  & dx/m & dy/m  & Lx/km   & Ly/km  \\
  \hline
1 & 105 & 105 & 2250 & 2250 & 236.25 & 236.25 \\
2 & 103 & 103 & 750  & 750  & 7725  & 77.25  \\
3 & 101 & 101 & 150  & 150  & 15.15  & 15.15  \\
4 & 201 & 201 & 30   & 30   & 6.03   & 6.03   \\
5 & 301 & 301 & 10   & 10   & 3.01   & 3.01   \\
\hline
\end{tabular*}
\label{Table:GridDesign}
\end{table*}

\subsection{Results and Discussion}

\begin{figure}[tbhp]
    \centering
    \includegraphics[width=\columnwidth]{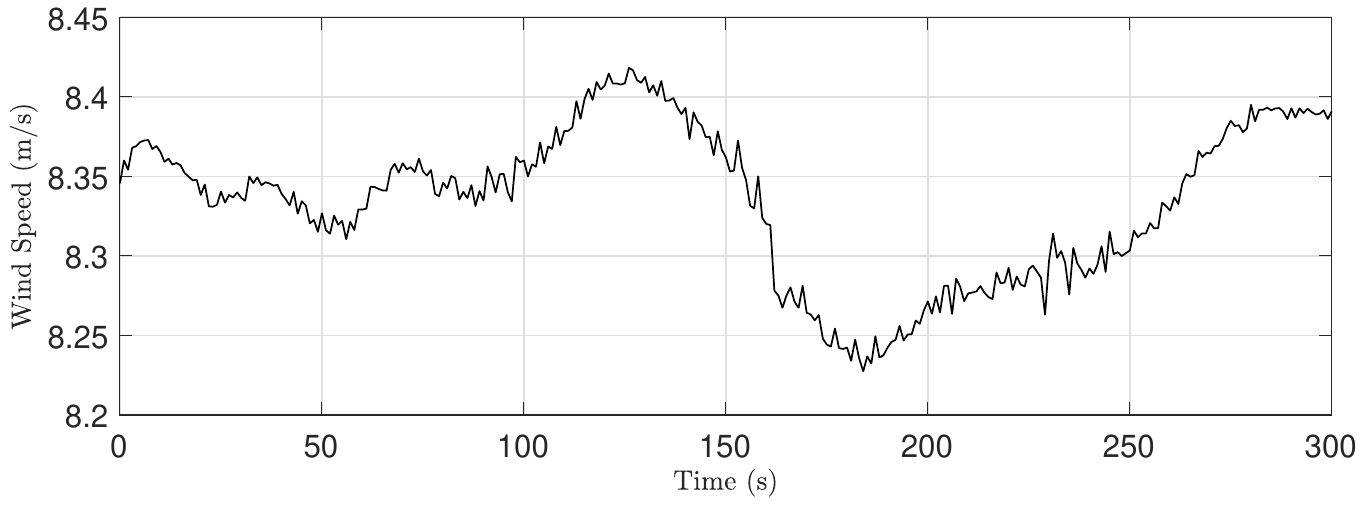}
    \caption{Wind speed at the hub height.}
    \label{fig:AverageVw}
\end{figure}

\begin{figure}[tbhp]
  \centering
  \subfigure[U component]{
  \begin{minipage}[t]{0.45\columnwidth}
  \centering
  \includegraphics[width=\linewidth]{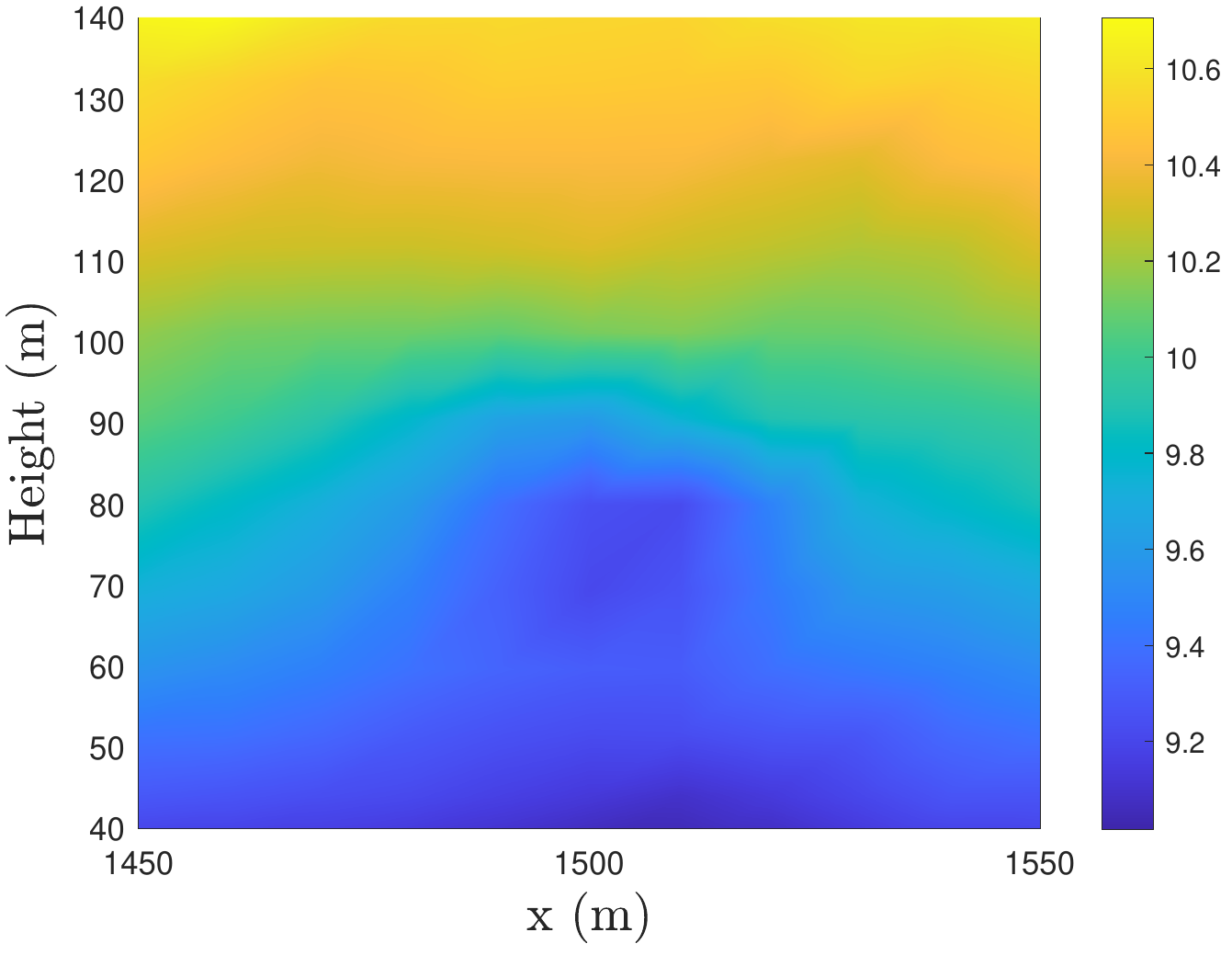}
  \label{fig:Fronta}
  \end{minipage}
  }
  \hfill
  \subfigure[V component]{
  \begin{minipage}[t]{0.45\columnwidth}
  \centering
  \includegraphics[width=\linewidth]{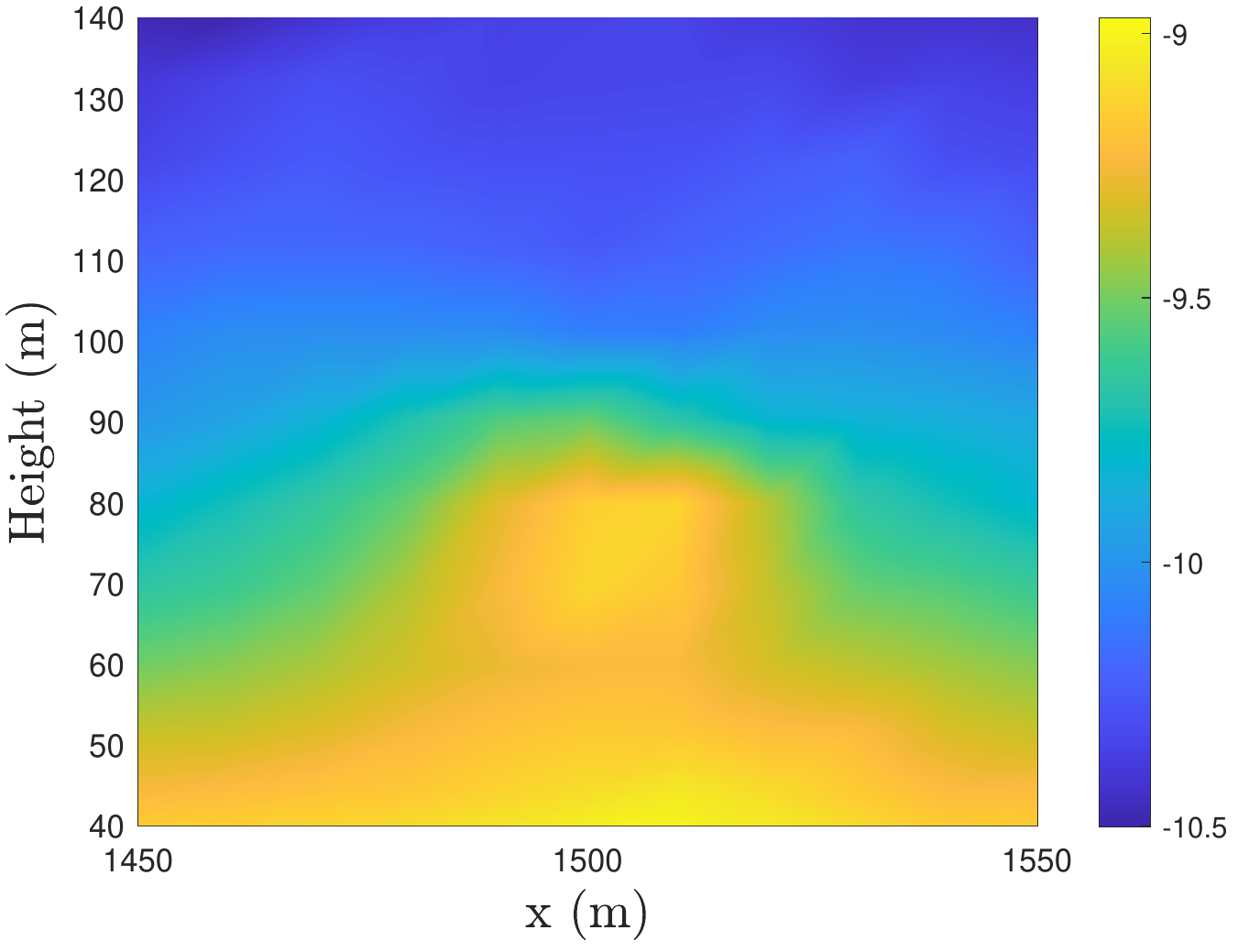}
  \label{fig:Frontb}
  \end{minipage}
  } 
  \vfill
    \subfigure[ W component]{
  \begin{minipage}[t]{0.45\columnwidth}
  \centering
  \includegraphics[width=\linewidth]{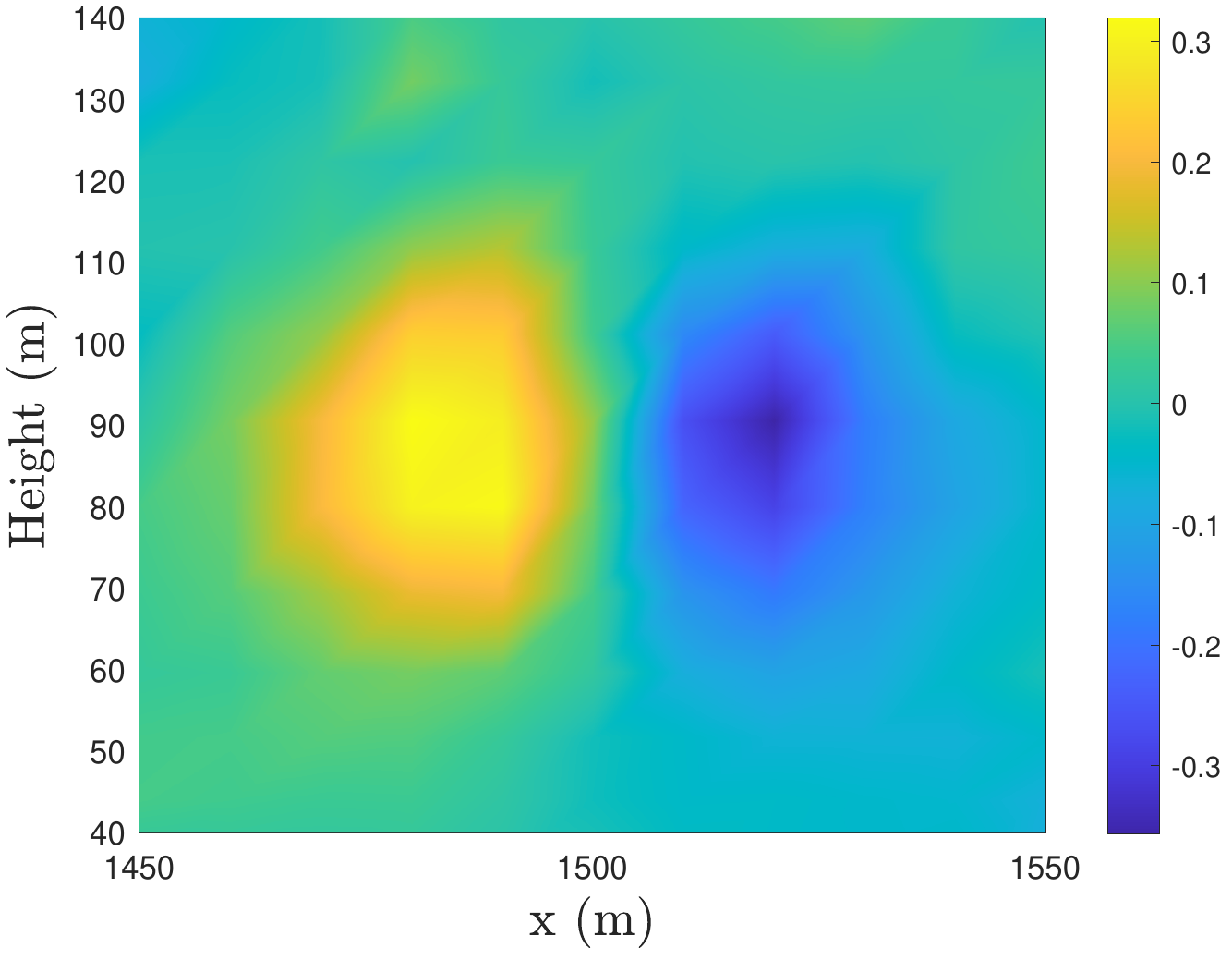}
  \label{fig:Frontc}
  \end{minipage}
  }
  \hfill
  \subfigure[Total wind speed]{
  \begin{minipage}[t]{0.45\columnwidth}
  \centering
  \includegraphics[width=\linewidth]{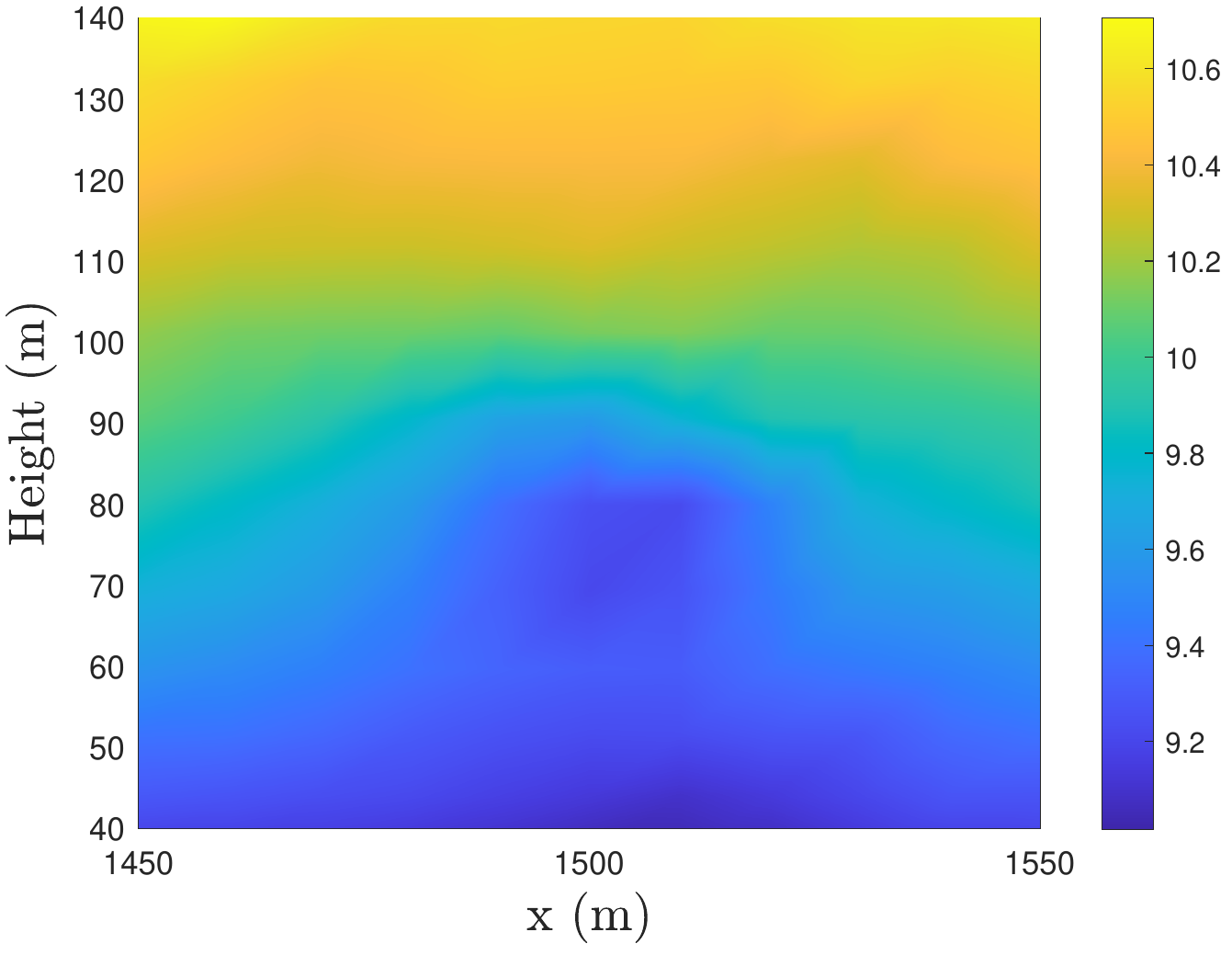}
  \label{fig:Frontd}
  \end{minipage}
  }  
  \caption{Wind field around the GAD region, at $t=90$ s.}
  \label{fig:Front}
\end{figure}

The wind speed at the hub height is presented in Fig.~\ref{fig:AverageVw}. To get a knowledge of the wind profile around the disc area, the wind field in the vertical plane around the turbine at $t = 90$ is presented in Fig.~\ref{fig:Front}. In Fig.~\ref{fig:Fronta}, the positive value refers to the north direction; In Fig.~\ref{fig:Frontb}, the positive value refers to the east direction;In Fig.~\ref{fig:Frontc}, the positive value refers to the upper direction. As shown in Fig.~\ref{fig:Front}, the wind speed varies in the overall disk district. Besides, the proposed WRF-IWT model is able to capture the wakes dynamics. In Fig.~\ref{fig:planeXY}, the horizontal view of wakes at hub height is presented. As shown in Fig.~\ref{fig:Front}-Fig.~\ref{fig:planeXY}, the proposed WRF-IWT model is able to capture the turbine's influence on the surrounding wind field, as well as the wake influence.

\begin{figure}
    \centering
    \includegraphics[width=\columnwidth]{figures/wakes_xy.pdf}
    \caption{Wind turbine wake at the hub height, in the horizontal plane, at $t=90$ s.}
    \label{fig:planeXY}
\end{figure}

   \begin{figure*}[tbhp]
  \centering
  \subfigure[Comparison of the rotor power $P_{\mathrm{r}}$.]{
  \centering
  \includegraphics[width=\columnwidth]{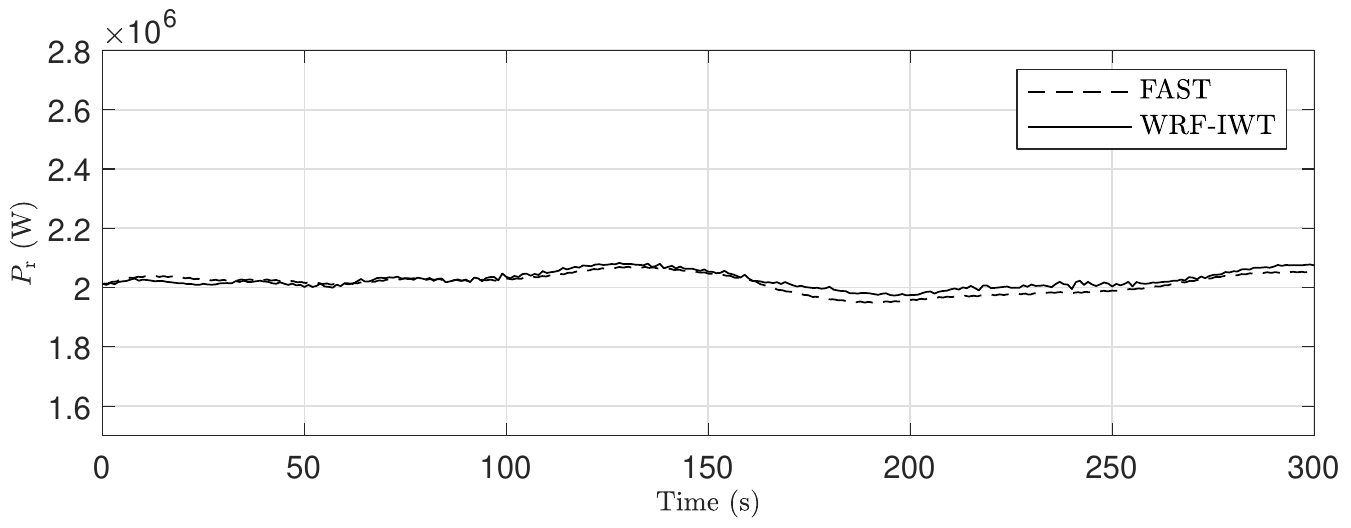}
  \label{fig:BTCa}
  }
  \hfill
  \subfigure[Comparison of the generator power $P_{\mathrm{g}}$.]{
  \centering
  \includegraphics[width=\columnwidth]{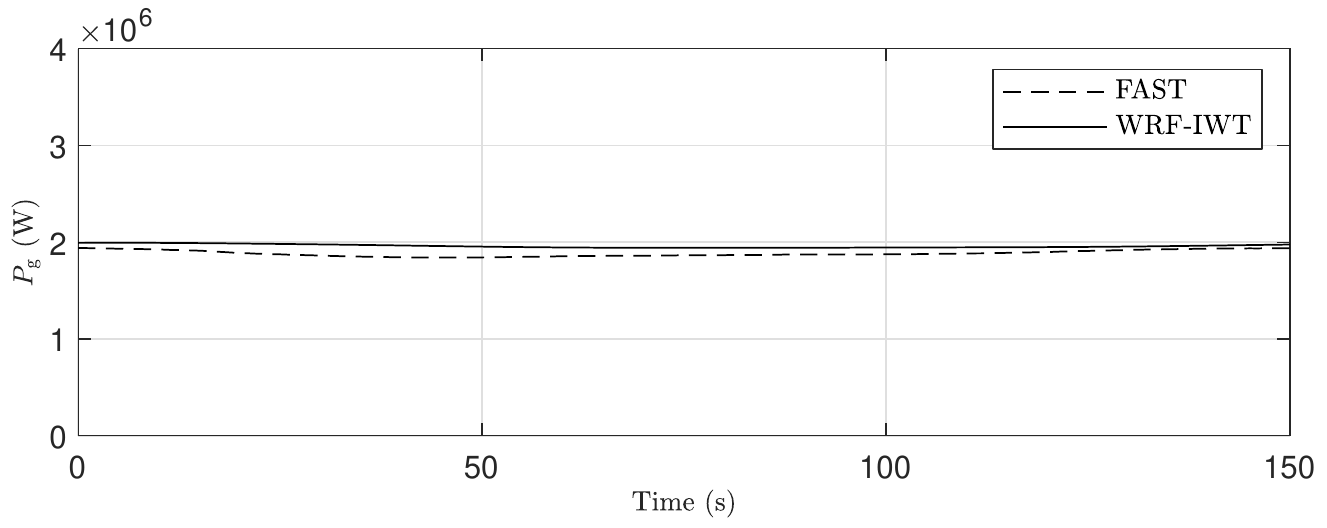}
  \label{fig:BTCb}
  }  \\
  \vfill
    \subfigure[Comparison of the generator torque $T_{\mathrm{g}}$.]{
    \centering
  \includegraphics[width=\columnwidth]{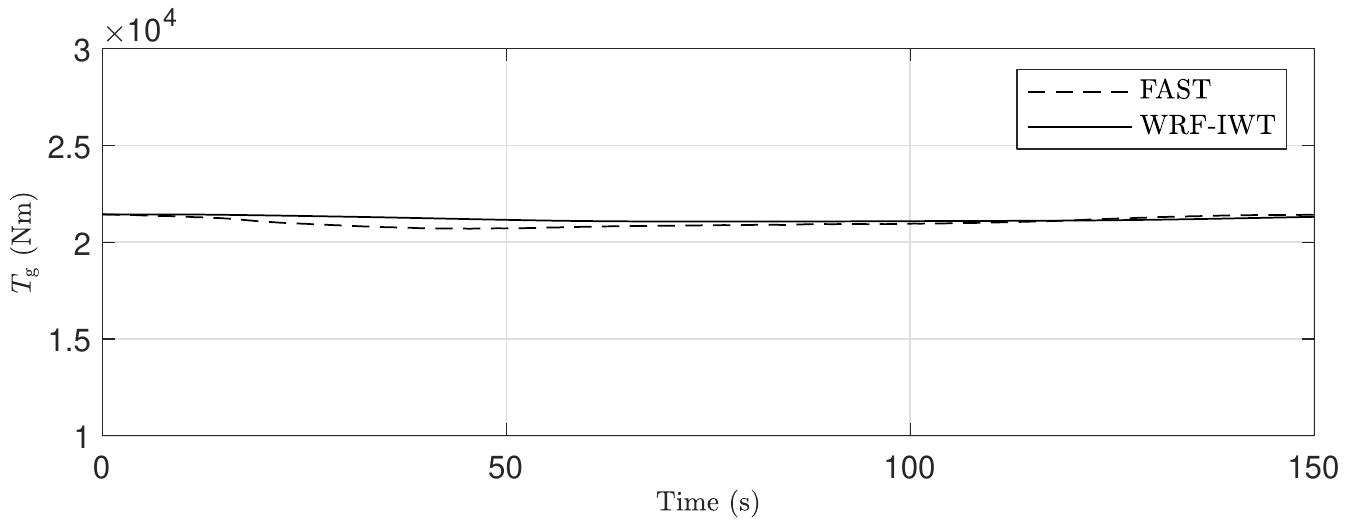}
  \label{fig:BTCc}
  }
  \hfill
  \subfigure[Comparison of the generator rotation speed $\omega_{\mathrm{g}}$.]{
  \centering
  \includegraphics[width=\columnwidth]{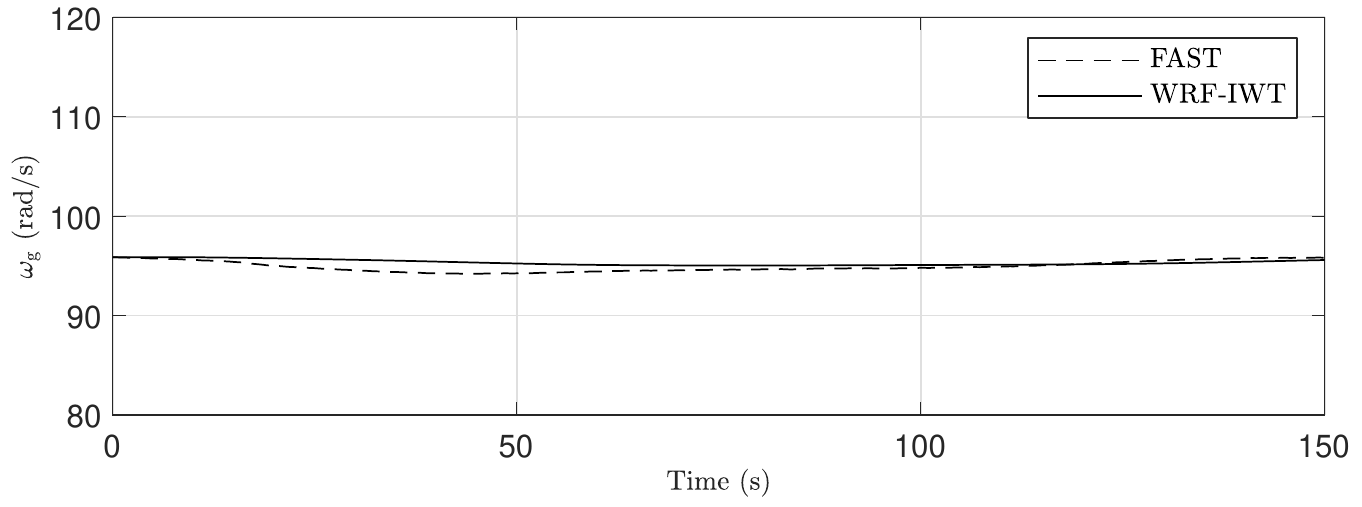}
  \label{fig:BTCd}
  }  
  \vfill
  \caption{Comparison of the control performance between the WRF-IWT framework and the FAST model.}
  \label{fig:BTC}
\end{figure*}

 In addition, the proposed WRF-IWT model has the capability to simulate the turbine dynamics and controller performance. Under the wind condition presented in Fig.~\ref{fig:AverageVw}, the wind speed is below the rated wind speed $11.4$ m/s. Thus, the BTC is activated~\shortcite{jonkman2009}. 
 In Fig.~\ref{fig:BTC}, the controller performance is presented and compared with results derived from FAST. 
 In Fig.~\ref{fig:BTCa}, the comparison of the captured rotor power $P_{\mathrm{r}}$ is presented. As shown in the plot, it matches quite well with the results derived by FAST. Besides, it is able to capture the time-varying variation of $P_{\mathrm{r}}$ caused by the local atmospheric perturbation and the tower vibration. Nevertheless, with the BTC this variation does not affect the smooth of generator power production $P_{\mathrm{g}}$, $\omega_{\mathrm{g}}$ and $T_{\mathrm{g}}$, as shown in Fig.~\ref{fig:BTCb}-Fig.~\ref{fig:BTCd}. Thus, the proposed WRF-IWT model is able to simulate the two-way coupling between the local atmospheric and the turbine with BTC locating inside that area. Besides, it also provide an insight of the BTC performance under the local atmospheric conditions with time-varying perturbation. At the same time, in general they also match the results derived by FAST. Because the proposed WRF-IWT model uses a simplified drive-train model. It is reasonable that the results do not match perfectly but have some minor differences. Therefore, the dynamics and controller in the proposed WRF-IWT model is validated to work properly.

\begin{figure}[tbhp]
    \centering
    \subfigure[The thrust force $T$ captured by the WRF-IWT model.]{
    \centering
    \includegraphics[width=\columnwidth]{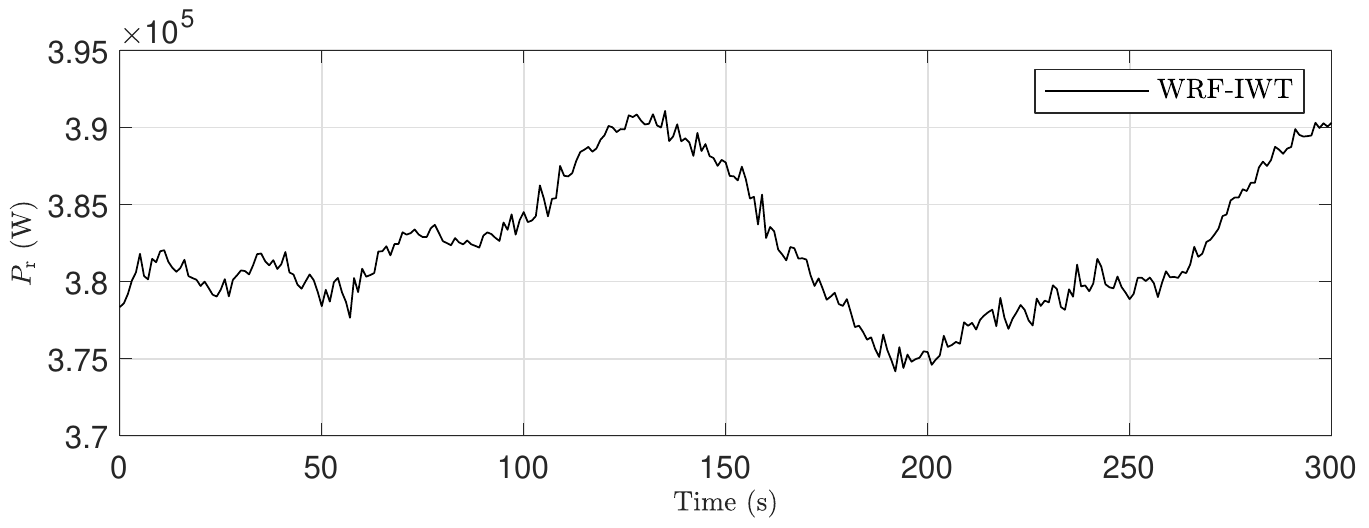}
    \label{fig:towera}
    }
    \vfill
     \subfigure[WRF-IWT: tower top velocity $V_{\mathrm{fa}}$.]{
    \includegraphics[width=\columnwidth]{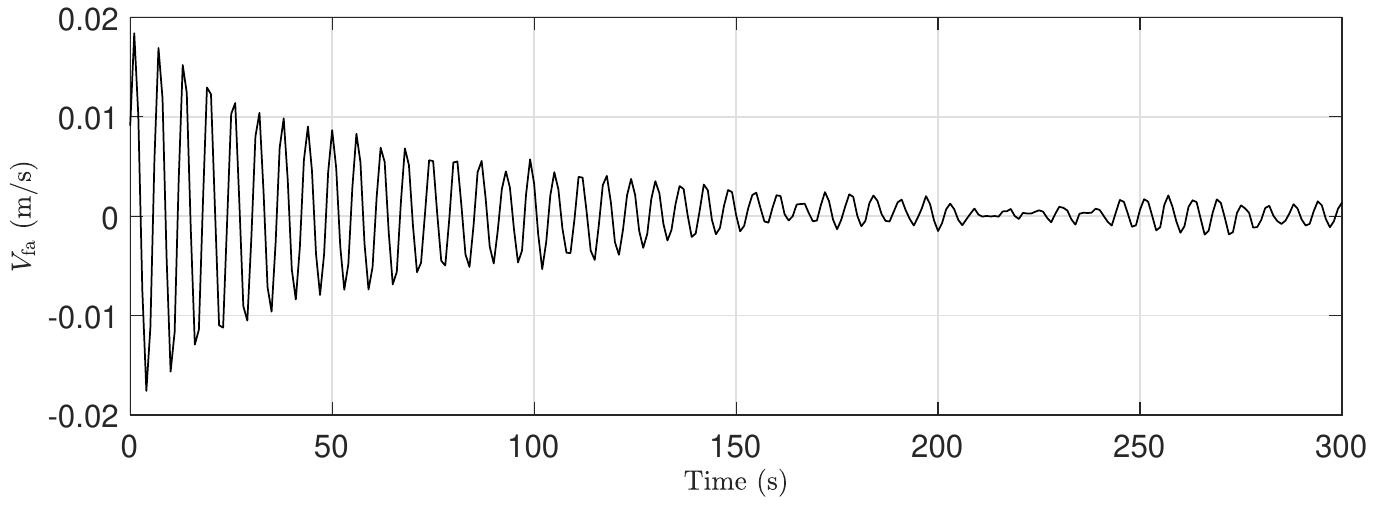}
    \label{fig:towerb}}
    \vfill
     \subfigure[WRF-IWT: tower top acceleration $A_{\mathrm{fa}}$.]{
    \includegraphics[width=\columnwidth]{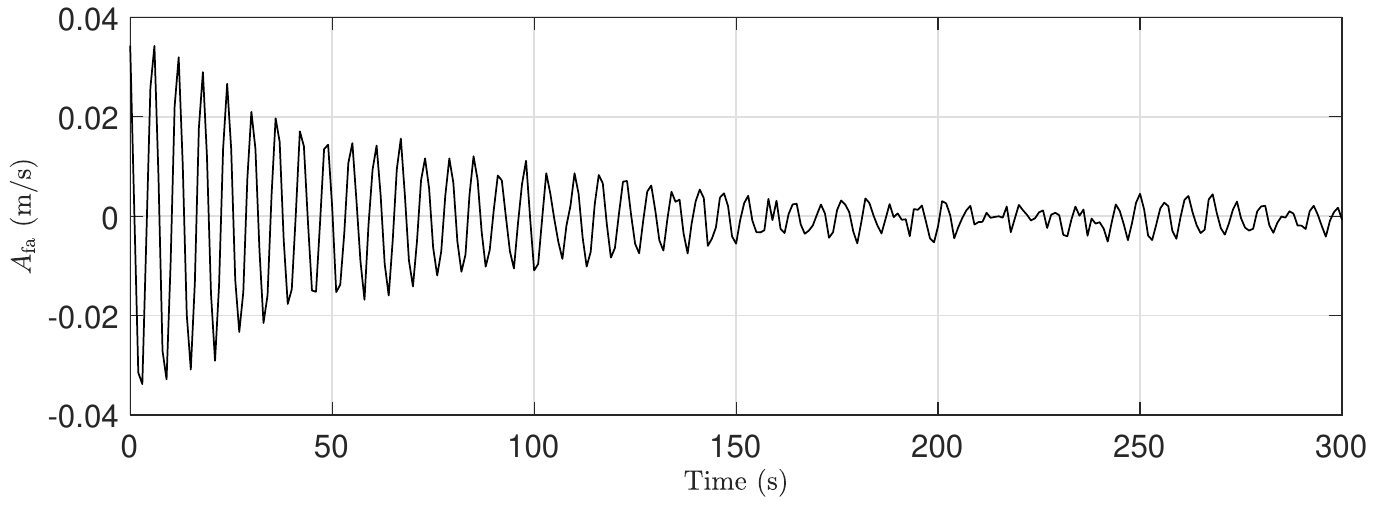}
    \label{fig:towerc}}
    \caption{Tower top vibration simulated via the WRF-IWT framework.}
    \label{fig:tower}
\end{figure}

Furthermore, based on the model in Section III.B, the proposed framework is able to simulate the tower vibration under short-term realistic local atmospheric conditions. The results are presented in Fig.\ref{fig:tower}, where $V_{\mathrm{fa}}$ and  $A_{\mathrm{fa}}$ are the tower top vibration velocity and acceleration, respectively. As shown in Fig.\ref{fig:towera}, the thrust force is influenced by the perturbation of local atmosphere and tower vibration, which shows a similar tendency as the $P_{\mathrm{r}}$. Besides, as shown in Fig.~\ref{fig:towerb}-Fig.~\ref{fig:towerc}, the tower vibration dynamics is also captured, which indicates that it is also influenced by the local wind perturbation. Thus, it is able to capture the tower vibration dynamics affected by the local atmospheric perturbation. This provides important information for further controller development with tower fatigue load mitigation.

In summary, the proposed WRF-IWT model is capable to capture the influence of a turbine with BTC on the local wind conditions, as well as the turbine dynamics influenced by the local atmospheric perturbation. Thus,it achieves to simulate the two-way coupling interaction between the local atmospheric conditions and a turbine with BTC. The simulation results not only provides an insight of the important information, but also show a potential of the proposed model to assist power optimization and tower fatige load reduction controller development.

\section{CONCLUSIONS}
In this paper, we develop a control-oriented wind turbine dynamic simulation framework considering time-varying local atmospheric conditions.
This is achieved by integrating an internal wind turbine (IWT) model in the weather research and forecast (WRF) simulation tool.
The IWT is composed of a generalized actuator disk (GAD) model, the drive-train dynamics and a tower top vibration model.
The GAD model and drive-train dynamics developed in WRF enables it to simulate the turbine dynamics and controller performance under short-term local atmospheric condition. 
By introducing the tower top vibration model, the proposed framework is able to capture the tower top motion and its effect on the relative wind speed between the rotor and the inflow wind speed. The proposed WRF-IWT framework is employed to verify the performance of a baseline torque controller (BTC) on the National Renewable Energy Laboratory (NREL) 5MW reference wind turbine situated near PLOCAN site. Simulation results illustrate that the WRF-IWT model is able to simulate the wind turbine and control dynamics where the local atmospheric condition around the PLOCAN site is considered. The results derived from the WRF-IWT model matches well with those derived from the Fatigue, Aerodynamics, Structures,and Turbulence (FAST) simulator.
Thus, the proposed framework shows high potential in assisting developing more applicable controllers for the wind turbine operating in the real world conditions, for both power optimization and tower fatigue load reduction objectives. Besides, in further research the current WRF-IWT framework has the potential to be extended to consider wakes effects for simulation of offshore wind turbines.


\section*{ACKNOWLEDGMENT}

This work is supported by both WATEREYE and HIPERWIND projects, which
are funded by European Innovation and Networks Executive
Agency under the European Union’s Horizon 2020 research
and innovation program under grant agreement No. 851207 and No. 101006689.

\bibliographystyle{chicaco}
\bibliography{references}

\end{document}